\documentclass[twoside]{article}
\usepackage{qic,epsfig}

\renewcommand{\thefootnote}{\fnsymbol{footnote}}  

\newcommand{\half}{\frac{1}{2}}
\newcommand{\trace}{\mathop{\rm Tr}\nolimits}
\newcommand{\re}{\mathop{\rm Re}\nolimits}

\newcommand{\diag}{\mathop{\rm Diag}\nolimits}

\newcommand{\be}{\begin{equation}}
\newcommand{\ee}{\end{equation}}
\newcommand{\bea}{\begin{eqnarray}}
\newcommand{\eea}{\end{eqnarray}}
\newcommand{\beas}{\begin{eqnarray*}}
\newcommand{\eeas}{\end{eqnarray*}}

\begin{document}
\setlength{\textheight}{8.0truein}    

\runninghead{Distinguishability measures}
            {Audenaert}

\normalsize\textlineskip
\thispagestyle{empty}
\setcounter{page}{1}

\vspace*{0.88truein}

\alphfootnote

\fpage{1}

\centerline{\bf
COMPARISONS BETWEEN QUANTUM STATE}
\vspace*{0.035truein}
\centerline{\bf DISTINGUISHABILITY MEASURES}
\vspace*{0.37truein}
\centerline{\footnotesize
KOENRAAD M.R.\ AUDENAERT}
\vspace*{0.015truein}
\centerline{\footnotesize\it Department of Mathematics, Royal Holloway, University
of London,}
\baselineskip=10pt
\centerline{\footnotesize\it Egham TW20 0EX, United Kingdom}
\vspace*{10pt}
\vspace*{0.225truein}

\vspace*{0.21truein}

\abstracts{
We provide a compendium of inequalities between several quantum state distinguishability measures.
For each measure these inequalities consist of the sharpest possible upper and lower bounds in terms of another measure.
Some of these inequalities are already known, but new or more general proofs are given, whereas other inequalities are new.
We also supply cases of equality to show that all inequalities are indeed the sharpest possible.
}{}{}

\vspace*{10pt}

\keywords{Trace norm distance, Uhlmann fidelity, relative entropy, Chernoff distance}
\vspace*{3pt}

\vspace*{1pt}\textlineskip    
\section{Introduction}
\setcounter{footnote}{0}
\renewcommand{\thefootnote}{\alph{footnote}}

Over the years, a large number of quantum state distinguishability measures have been defined and subsequently studied.
Prime examples are the trace norm distance, Uhlmann fidelity, relative entropy and Chernoff distance.
Each of these measures has different properties, hence in different situations one or the other measure may be most
suitable, or practical. It is therefore important to know how the different measures are related to one another.
In particular, inequalities providing bounds on a measure in terms of another measure are very useful.
One well-known such `sandwich' bound is the one which says that the Uhlmann fidelity $F$
is bounded above by $\sqrt{1-T^2}$ and bounded below by $1-T$,
where $T$ is the trace norm distance (see below for the exact definitions).
Is this bound best possible? And what about comparisons of $F$ with other measures?

In this note we present best possible dimension-independent
upper and lower bounds (where they exist) between the following well-established
distinguishability measures:
relative entropy, Chernoff distance,  Uhlmann fidelity, trace norm distance and
the linear overlap (admittedly, this one is only useful for pure states). In addition,
we also include a quantity which is essentially the Renyi relative entropy with parameter $1/2$,
henceforth denoted by the letter $Q$.
The reason for including this little-used measure is mathematical convenience, although it may deserve more
prominence than it currently enjoys.

With 6 measures, in principle one has to consider 30 pairs (even though not every pair yields an inequality;
for example, there are no upper bounds on the relative entropy in terms of the other measures).
Fortunately, most combinations can be derived from a handful of essential inequalities simply by chaining two or
more of them together.

After introducing some basic mathematical definitions and tools, we give the definitions of the distinguishability measures
that we study in this note in Section \ref{sec:distances}. The essential inequalities are given in
Section \ref{sec:ess} together with short proofs. Some of these inequalities have appeared already in the
literature; nevertheless, the proofs given here are new unless indicated.
Finally, in Section \ref{sec:ineq} we list all possible lower and upper bounds on each of the distinguishability measures
considered, together with a number of cases of equality. All bounds provided are dimension-independent, and they are all
best possible, as attested by the cases of equality.
We hope that this compendium of relations will be found useful by the quantum information community.
\section{Preliminaries}
We will denote a diagonal matrix with diagonal elements $x_1,\ldots,x_n$ by $X=\diag(x_1,\ldots,x_n)$.
We will use the convention $0^t:=0$ for all real $t$ and $0\log 0=0$.
That is, for positive semidefinite $A$, $A^t$ and $A\log A$ are defined to be zero outside the support of $A$.
For any matrix $X$ let $X^*$ denote its Hermitian conjugate and $|X|$ its absolute value (modulus); $|X|:=\sqrt{X^*X}$.
The Jordan decomposition of a Hermitian matrix $X$ into its positive and negative parts is given by
$X=X_+ - X_-$, with $X_\pm = (|X|\pm X)/2$, and $|X|=X_++X_-$.

The Schatten $q$-norms are the non-commutative generalisations of the $\ell_q$ norms. For any matrix $A$,
its $q$-norm $||A||_q$, for $1\le q$, is defined as
$$
||A||_q = \left(\trace|A|^q\right)^{1/q}.
$$
One sees that this norm is equal to the $\ell_q$ norm of the vector of singular values of $A$.
The Schatten 1-norm ($q=1$) is well-known in quantum information theory
as the trace norm: $||A||_1 = \trace |A|$.
It is equal to the sum of the singular values of $A$. In terms of the positive and negative parts of $A$,
$||A||_1 = \trace A_+ + \trace A_- = 2\trace A_+ - \trace A$.
Basic properties of the $q$-norm are:
if $p<q$ then $||A||_q\le ||A||_p$; for positive semidefinite $A$ and $p>0$, $||A^p||_q = ||A||^p_{pq}$.
We will occasionally need the quantity $||A||_q$ for values of $q$ less than 1; while the definition allows this, it has to be
kept in mind that this is no longer a norm but a quasi-norm.

An important inequality involving the Schatten norms is H\"older's inequality:
for all matrices $A$, $B$, and indices $p$, $q$, $r$ satisfying $p,q,r\ge 1$ and $1/p+1/q\le 1/r$,
$$
||AB||_r \le ||A||_p\;\;||B||_q.
$$
Extensions to products of more than 2 factors are obvious.
More information can be found, e.g.\ in the standard textbook \cite{bhatia}.
\section{Distinguishability measures for pairs of states\label{sec:distances}}
Let $\rho$ and $\sigma$ be two density matrices.
The following quantities are distinguishability measures between the two states,
all of which (except $L$) have the property that they
achieve their extremal (minimal or maximal) values for $\rho=\sigma$ and for $\rho\sigma=0$, respectively.
\beas
L(\rho,\sigma) &=& \trace\rho\sigma \\
T(\rho,\sigma) &=& ||\rho-\sigma||_1 /2\\
F(\rho,\sigma) &=& ||\rho^{1/2}\sigma^{1/2}||_1\\
Q_{s}(\rho,\sigma) &=& \trace(\rho^s\sigma^{1-s})\\
Q(\rho,\sigma) &=& Q_{1/2}(\rho,\sigma)\\
Q_{\min}(\rho,\sigma) &=& \min_{0\le s\le 1} Q_s(\rho,\sigma) \\
C(\rho,\sigma) &=& -\log Q_{\min}(\rho,\sigma)\\
S(\rho||\sigma) &=& \trace\rho(\log\rho-\log\sigma).
\eeas
Here,
$L$ is known as the linear fidelity, or overlap,
$T$ as the trace norm distance, $F$ as the Uhlmann fidelity, $Q_s$ (essentially) as the Renyi relative
entropy of order $s$, $C$ as the Chernoff distance and $S$ as the relative entropy.
It has to be kept in mind that the relative entropy $S(\rho||\sigma)$ becomes infinite whenever the support
of $\sigma$ is not contained in the support of $\rho$; in that case any upper bound on any of the distinguishability
measures in terms of the relative entropy becomes trivial.
For commuting $\rho$ and $\sigma$, the quantities $Q$ and $F$ coincide; they are two different non-commutative
generalisations of a quantity that is alternatively
known as the Hellinger affinity or the Bhattacharyya coefficient.

Note that $L$, $T$, $F$, $Q_{\min}$ and $C$ are symmetric in their arguments; that is,
the order of $\rho$ and $\sigma$ does not matter. In the case of $F$,
this is because $\rho$ and $\sigma$ are Hermitian; for
Hermitian $A$ and $B$ and any unitarily invariant norm,  $|||AB|||=|||BA|||$.

The definitions of these quantities can be extended to non-normalised states, i.e.\ positive semidefinite matrices,
allowing to absorb prior probabilities into the states.
In what follows, let $p$ be the prior probability of $\rho$ and $1-p$ the one of $\sigma$, and let $A=p\rho$
and $B=(1-p)\sigma$ so that $A,B\ge0$ and $\trace(A+B)=1$.
We redefine the above distinguishability measures in terms of $A$ and $B$ in such a way that setting $p=1/2$ we recover the
original definitions for normalised states:
\beas
L(A,B) &=& 4\trace AB \\
T(A,B) &=& ||A-B||_1 \\
F(A,B) &=& 2||A^{1/2}B^{1/2}||_1\\
Q_s(A,B) &=& 2\trace(A^{s}B^{1-s})\\
Q(A,B) &=& 2\trace(A^{1/2}B^{1/2}).
\eeas

For every pair of these quantities one can ask for the sharpest possible
upper and lower bounds on one quantity in terms
of the other. Of the many possible combinations only a few inequalities turn out to be essential;
by chaining them together in various ways all other combinations can easily be obtained.
\bea
L(A,B) &\le& F^2(A,B) \label{eq:LF} \\
1-Q_s(A,B) &\le& T(A,B)  \label{eq:QT} \\
T^2(A,B) &\le& 1-F^2(A,B) \label{eq:TF} \\
F^2(\rho,\sigma) &\le& Q_s(\rho,\sigma), \forall 0\le s\le 1 \label{eq:FQs} \\
Q(A,B) &\le& F(A,B) \label{eq:QF}.
\eea
For example, chaining $T^2\le 1-F^2$ with $Q\le F$ yields $T^2\le 1-Q^2$,
and chaining $1-Q\le T$ with $Q\le F$ yields $1-T\le F$, neither of which can be improved.
On the other hand, chaining $1-Q\le T$ with $T^2\le 1-F^2$ yields $1-\sqrt{1-F^2}\le Q$,
but this is weaker than $F^2\le Q$ (which follows immediately from (\ref{eq:FQs})).
We will present all optimal chainings in Section \ref{sec:ineq}, together with cases of equality.
\section{Essential inequalities, with proofs\label{sec:ess}}
We now present  the essential inequalities as theorems, and provide short proofs.
\begin{theorem}
For $0\le s\le 1$, $1-T(A,B)\le Q_s(A,B)$.
\end{theorem}
This inequality was first proven for all $s\in[0,1]$ in \cite{ANSzV},
although the special case for $s=1/2$ had already been proven much earlier by Powers and St{\o}rmer \cite{powers},
with a generalisation to von Neumann algebras by Araki \cite{araki}.
In these works, the inequality appears in the form $||A^\half-B^\half||_2^2 \le ||A-B||_1$.
The book \cite{hayashi} by Hayashi also contains a proof of the $s=1/2$ case.
A much shorter proof of (\ref{eq:QT}), due to N.~Ozawa, appeared recently in \cite{jaksic},
and allowed generalisation of this inequality to von Neumann algebras \cite{ogata}.

\noindent\textbf{Proof:} (Ozawa).
Let $\Delta=A-B$ and define $C=B+\Delta_+=A+\Delta_-$.
Clearly, $0\le A,B\le C$. Hence, by operator monotonicity of the fractional power $x^s$ (for $0\le s\le 1$),
$0\le A^s,B^s\le C^s$ and $0\le A^{1-s},B^{1-s}\le C^{1-s}$.
Therefore, we have the chain of inequalities
\beas
\trace(A-B^{1-s}A^{s})
&=& \trace(A^{1-s}-B^{1-s})A^{s} \\
&\le& \trace(C^{1-s}-B^{1-s})A^{s} \\
&\le& \trace(C^{1-s}-B^{1-s})C^{s} \\
&=& \trace(C-B^{1-s}C^{s}) \\
&=&\trace\Delta_+ + \trace(B-B^{1-s}C^{s}) \\
&=& \trace\Delta_+ + \trace B^{1-s}(B^{s}-C^{s}) \\
&\le& \trace\Delta_+.
\eeas
Thus, $1-T(A,B) = 2\trace(A-\Delta_+) \le 2\trace B^{1-s}A^{s}=Q_{s}(A,B)$.
\square

\begin{theorem}
$L(A,B)\le F^2(A,B)$.
\end{theorem}
\textbf{Proof:}
This follows from the inequality $||X||_1\le ||X||_{1/2}$ applied to $X=B^{1/2}AB^{1/2}$
and noting that $\trace(B^{1/2}AB^{1/2})=\trace(AB)=L(A,B)/4$ and
$||B^{1/2}AB^{1/2}||_{1/2}=||A^{1/2}B^{1/2}||_1^2=F^2(A,B)/4$.
\square

\noindent
Inequalities (\ref{eq:FQs}) and (\ref{eq:QF}) also appeared in \cite{ANSzV} (its Theorem 6 and equation (28), respectively).

\begin{theorem}
For $0\le s\le 1$,
$F(\rho,\sigma) \le (\trace\rho^{s} \sigma^{1-s})^{1/2}$.
\end{theorem}
\textbf{Proof:}
We rewrite $\rho^{1/2} \sigma^{1/2}$ as a product of three factors
$$
\rho^{1/2} \sigma^{1/2} = \rho^{(1-s)/2} (\rho^{s/2}\sigma^{(1-s)/2}) \sigma^{s/2},
$$
apply H\"older's inequality on the trace norm of this product, and exploit the
relation $\|X^p\|_q = \|X\|^p_{pq}$
(for $X\ge0$):
\beas
\|\rho^{1/2} \sigma^{1/2}\|_1 &=& \|\rho^{(1-s)/2} (\rho^{s/2}\sigma^{(1-s)/2}) \sigma^{s/2}\|_1 \\
&\le& \|\rho^{(1-s)/2}\|_{2/(1-s)} \, \|\rho^{s/2}\sigma^{(1-s)/2}\|_{2}\, \|\sigma^{s/2}\|_{2/s} \\
&=& (\trace\rho)^{(1-s)/2} \, \|\rho^{s/2}\sigma^{(1-s)/2}\|_{2}\, (\trace\sigma)^{s/2} \\
&=& (\trace\rho^{s}\sigma^{1-s})^{1/2}.
\eeas
\square

\begin{theorem}
$Q(A,B)\le F(A,B)$.
\end{theorem}
\textbf{Proof:}
This follows from the inequality $|\trace X|\le||X||_1$
applied to $X=A^{1/2}B^{1/2}$, which has positive trace, hence $|\trace X|=\trace X$.
\square

\begin{theorem}
$F^2(A,B)+T^2(A,B)\le 1$.
\end{theorem}
This inequality was first proven by Fuchs and van de Graaf \cite{fuchs}, albeit only for a uniform prior. We provide a general
and shorter proof; again this proof can be generalised to the von Neumann algebra setting \cite{araki}.

\noindent\textbf{Proof:}
First note that, for any unitary matrix $U$, we have
$$
2(A-B) = (A^\half+UB^\half)^*(A^\half-UB^\half) + (A^\half-UB^\half)^*(A^\half+UB^\half).
$$
Exploiting first the triangle inequality for the trace norm, and then the Cauchy-Schwarz inequality yields
$$
||A-B||_1 \le ||A^\half+UB^\half||_2\;||A^\half-UB^\half||_2.
$$
Taking the square and simplifying the right hand side then gives
$$
T(A,B)^2 = ||A-B||_1^2 \le (\trace A+\trace B)^2 - (2\re\trace(A^\half UB^\half))^2 = 1 - (2\re\trace(A^\half UB^\half))^2.
$$
As this holds for any unitary matrix $U$, the sharpest bound is obtained by minimising over all $U$.
Noting that $\max_U \re\trace(A^\half UB^\half) = ||A^\half B^\half||_1 = F(A,B)$ yields the inequality of the theorem.
\square

\begin{theorem}
$-\log Q(\rho,\sigma)\le\half S(\rho||\sigma)$.
\end{theorem}
\textbf{Proof:}
By Lemma 2.1 in \cite{finiten}, the function $s\mapsto \psi(s):=\log\trace\rho^s\sigma^{1-s}$
is convex over $[0,1]$, for any choice of states $\rho$ and $\sigma$.
This implies the following inequality:
$$
\frac{\psi(1)-\psi(1/2)}{1-1/2} \le \psi'(1).
$$
Since $\psi(1)=0$, $\psi(1/2)=\log Q(\rho,\sigma)$ and $\psi'(1) = S(\rho||\sigma)$, we immediately get
the inequality of the Theorem.
\square

The sharpest lower bound on the relative entropy in terms of the trace norm distance was found by
Hiai, Ohya and Tsukuda \cite{hot}:
\begin{theorem}[H-O-T]
$S(\rho||\sigma) \ge s(||\rho-\sigma||_1 / 2)$.
\end{theorem}
Here, $s(x)$ is a special function defined for $0\le x< 1$ as
\beas
s(x) &:=& \min_{0<r<1-x} S(\diag(r+x,1-r-x)||\diag(r,1-r))\\
&=& \min_{x<r<1} S(\diag(r-x,1-r+x)||\diag(r,1-r)).
\eeas
In \cite{baka2},
the first three non-zero terms in its series expansion around $x=0$ were determined:
\begin{equation}\nonumber
s(x) =
2 x^2 + \frac{4}{9}x^4 + \frac{32}{135}x^6 + O(x^{8}). \label{bound_ohya_better}
\end{equation}
One can easily prove \cite{csiszar} that
the lowest order expansion $2x^2$ is actually a lower bound, known as Pinsker's bound.
For values of $x>4/5$,
$s(x)$ is well approximated by its upper bound
\begin{eqnarray*}
    s(x) &\le & \lim_{r\rightarrow 1-x} S(\diag(r+x,1-r-x)||\diag(r,1-r))\nonumber\\
    & = & -\log(1-x).
\end{eqnarray*}

\section{All inequalities, with cases of equality\label{sec:ineq}}
In the following, we consider four special families of pairs of states.
\begin{itemlist}
\item
\textbf{Family (a):}
$$
\rho=\diag(1,0),\quad \sigma=\diag(t,1-t), 0\le t\le 1.
$$
For these states we have
$$
L=F^2=Q_{\min}=Q^2=1-T=\exp(-S)=t.
$$

\item
\textbf{Family (b):} Both $\rho$ and $\sigma$ are pure.
For these states we have
$$
L=F^2=Q_{\min}=Q=1-T^2.
$$
The relative entropy is infinite except for $\rho=\sigma$, in which case it is 0.

\item
\textbf{Family (c):}
$$
\rho=\diag(1-t,t,0),\quad \sigma=\diag(1-t,0,t), 0\le t\le 1.
$$
For these states we have
$$
\sqrt{L}=F=Q_{\min}=Q=1-T=1-t.
$$
The relative entropy is infinite except for $t=0$, in which case it is 0.

\item
\textbf{Family (d):}
$$
\rho=\diag(1-t,t),\quad \sigma=\diag(t,1-t), 0\le t\le 1.
$$
For these states we have
$$
F=Q_{\min}=Q=2\sqrt{t(1-t)}=\sqrt{2L};\quad T=|1-2t|=\sqrt{1-F^2}.
$$
The relative entropy is given by $S=(2t-1)\log(t/(1-t))$.
\end{itemlist}

The basic inequalities that were proven in the above can be chained together in various ways.
Below we present, for each of the considered distinguishability measures, the optimal lower and upper bounds
(if there are any) in terms of the other measures. The letters above the inequality signs indicate for which of the
four families of states considered above equality is obtained.

We do not attempt to provide all cases of equality;
the goal here is merely to show that the presented inequalities cannot be improved. For example,
the fact that all inequalities in the chain $L\le F^2\le Q_{\min} \le Q$ become equalities for the same family of states
(in this case family (b)) shows that $L\le Q$ is the best possible upper bound on $L$ in terms of $Q$.

To shorten the formulas, we define the function
$$
v(x) := \sqrt{1-x^2}.
$$
\begin{itemlist}
\item\textbf{$L$:}
$$
L \stackrel{a,b,c}{\le} F^2 \quad\left\{
\begin{array}{l}
\stackrel{a,b}{\le} Q_{\min} \stackrel{b,c,d}{\le} Q \\[3mm]
\stackrel{b,d}{\le} 1-T^2
\end{array}
\right.
$$
There are no lower bounds on $L$ in terms of the other measures because $L$ is unable to
indicate equality of the states when not both of them are pure; indeed, by taking $\rho=\sigma$ every other measure
would give 0 (or 1), but $L$ could assume any value between 0 and 1 depending on the purity of $\rho$.

There is no upper bound on $L$ in terms of $S$. For the choice $\rho=\diag(1-\epsilon,\epsilon)$ and
$\sigma=\diag(1,0)$, $L=1-\epsilon$ and $S=\infty$, whereas for the choice
$\rho=\sigma=\diag(1-x,x)$ with $x=\epsilon/2$, $L\approx 1-\epsilon$ too but $S=0$.
\item\textbf{$F$:}
$$
\left.
\begin{array}{r}
\left.
\begin{array}{r}
1-T \stackrel{c}{\le} Q_{\min} \stackrel{b,c,d}{\le} \\ [3mm]
\exp(-S/2) \stackrel{a}{\le}
\end{array}
\right\}\quad
Q \stackrel{a,c,d}{\le} \\ [3mm]
\sqrt{L} \stackrel{a,b,c}{\le}
\end{array}
\right\}\quad F
\quad\left\{
\begin{array}{l}
\stackrel{a,b}{\le} \sqrt{Q_{\min}} \stackrel{b,c,d}{\le} \sqrt{Q} \\ [3mm]
\stackrel{b,d}{\le} v(T)
\end{array}
\right.
$$
From this diagram one obtains, for example, the known sandwich bound $1-T\le F\le \sqrt{1-T^2}$.
There is no upper bound on $F$ in terms of $S$, as it would imply an upper bound on $S$ in terms of $F$;
however, $S$ can be infinite for any allowed value of $F$.
\item\textbf{$T$:}
$$
\left.
\begin{array}{r}
1-F \stackrel{a,c,d}{\le} 1-Q \stackrel{b,c,d}{\le} 1-Q_{\min} \stackrel{a,c}{\le} \\ [3mm]
s^{-1}(S) \le
\end{array}
\right\} \quad
T \quad
\left\{
\begin{array}{l}
\stackrel{b,d}{\le} v(F) \stackrel{a,c,d}{\le} v(Q) \stackrel{b,c,d}{\le} v(Q_{\min}) \\ [3mm]
\stackrel{a}{\le} \sqrt{1-L}
\end{array}
\right.
$$
Equality in the H-O-T bound is achieved for certain $2\times 2$ diagonal states.
\item\textbf{$Q$:}
$$
\left.
\begin{array}{r}
\left.
\begin{array}{r}
L \stackrel{a,b,c}{\le} F^2 \stackrel{a,b}{\le} \\ [3mm]
1-T \stackrel{c}{\le}
\end{array}
\right\} \quad
Q_{\min} \stackrel{b,c,d}{\le} \\ [3mm]
\exp(-S/2) \stackrel{a}{\le}
\end{array}
\right\}
\quad
Q
\stackrel{a,c,d}{\le} F
\quad \left\{
\begin{array}{l}
\stackrel{a,b}{\le} \sqrt{Q_{\min}} \\ [3mm]
\stackrel{b,d}{\le} v(T)
\end{array}
\right.
$$
\item\textbf{$Q_{\min}$:}
$$
\left.
\begin{array}{r}
\left.
\begin{array}{r}
L \stackrel{a,b,c}{\le} \\ [3mm]
\exp(-S) \stackrel{a}{\le} Q^2 \stackrel{a,c,d}{\le}
\end{array}
\right\}
\quad F^2 \stackrel{a,b}{\le} \\ [3mm]
1-T \stackrel{a,c}{\le}
\end{array}
\right\} \quad
Q_{\min} \stackrel{b,c,d}{\le} Q \stackrel{a,c,d}{\le} F \stackrel{b,d}{\le} v(T)
$$
Noting that the Chernoff divergence $C$ is just minus the logarithm of $Q_{\min}$ we immediately obtain bounds on $C$.
For example: $C\le S$ and $C\le -2\log F \le -2\log Q\le 2C$.
\end{itemlist}

\nonumsection{Acknowledgements}
\noindent
Thanks to Matthias Christandl for his encouragement and for raising the question how the measures $Q$ and $S$ are related.

\nonumsection{References}

\end{document}